\documentclass[aps,pra,twocolumn,nobalancelastpage]{revtex4-1}
\usepackage[latin9]{inputenc}
\usepackage{color}
\usepackage{bm}
\usepackage{amsmath}
\usepackage{amssymb}
\usepackage{graphicx}
\usepackage{esint}
\usepackage[colorinlistoftodos]{todonotes}
\usepackage{notoccite}
\usepackage[unicode=true,pdfusetitle,
 bookmarks=false,
 breaklinks=true,pdfborder={0 0 0},backref=false,colorlinks=true]
 {hyperref}
\usepackage[caption=false]{subfig}
\hypersetup{linkcolor=blue, urlcolor=blue, citecolor=blue}

\graphicspath{ {main_figures/} {supplement_figures/} }

\begin{document}

\title{Quarter-Flux Hofstadter Lattice in Qubit-Compatible Microwave Cavity Array}

\author{Clai Owens}
\author{Aman LaChapelle}
\author{Brendan Saxberg}
\author{Brandon M. Anderson}
\author{Ruichao Ma}
\author{Jonathan Simon}
\author{David I. Schuster}

\affiliation{James Franck Institute and the University of Chicago}
\begin{abstract}

Topological and strongly correlated materials are exciting frontiers in condensed matter physics, married  prominently in studies of the fractional quantum hall effect~\cite{RevModPhys.71.S298}. There is an active effort to develop synthetic materials where the microscopic dynamics and ordering arising from the interplay of topology and interaction may be directly explored. In this work we demonstrate a novel architecture for exploration of topological matter constructed from tunnel-coupled, time-reversal-broken microwave cavities that are both low loss and compatible with Josephson junction-mediated interactions~\cite{wallraff2004strong}. Following our proposed protocol~\cite{anderson2016engineering} we implement a square lattice Hofstadter model at a quarter flux per plaquette ($\alpha=1/4$), with time-reversal symmetry broken through the chiral Wannier-orbital of resonators coupled to Yttrium-Iron-Garnet spheres. We demonstrate site-resolved spectroscopy of the lattice, time-resolved dynamics of its edge channels, and a direct measurement of the dispersion of the edge channels. Finally, we demonstrate the flexibility of the approach by erecting a tunnel barrier and investigating dynamics across it. With the introduction of Josephson junctions to mediate interactions between photons, this platform is poised to explore strongly correlated topological quantum science for the first time in a synthetic system.

\end{abstract}
\maketitle

\section{Introduction}

Initial interest in the quantum Hall effect arose from the unexpected observation of quantized transport in electronic heterostructures~\cite{klitzing1980new}. In the intervening years, we have come to understand that this property is the direct result of a topological winding number of the Bloch wave function in the Brillioun zone: physics which is robust not only to disorder but also to inter-particle collisions~\cite{RevModPhys.71.S298}. In this interacting case, the lowest-lying excitations, ``anyons,'' are of particular interest, as they are quasi-particles of fractional charge~\cite{de1998direct}, believed to have fractional statistics~\cite{stern2008anyons}. To date, definitive proof of  fractional statistics remain elusive, though transport measurements in anyon interferometers~\cite{an2011braiding,willett2010alternation} are highly suggestive~\cite{PhysRevB.80.155303}.

With new material platforms come new measurement techniques and new perspectives on the underlying physics; both particle-by-particle construction of topological fluids ~\cite{umucalilar2013many} and impurity interferometry ~\cite{grusdt2016interferometric} promise direct experimental signatures of the geometric phase acquired when anyons are transported around one another, but require either the ability to construct a small Laughlin puddle one particle at a time, or the binding of an anyon to a mobile impurity that is itself transported through an interferometer. While it is unclear how to achieve this in electronic materials, synthetic material platforms have begun to emerge where such microscopic control is feasible.

Synthetic topological materials fall into two principal categories: those made of ultracold atoms~\cite{bloch2008many} and those made of light. In both cases the challenges are (1) to engineer a synthetic gauge field for the (charge neutral) particles; and (2) to mediate interactions between them. In the case of ultracold atoms, s-wave contact interactions arise naturally~\cite{greiner2002quantum}, while inducing gauge fields requires rotation of the atomic gas~\cite{doi:10.1080/00018730802564122,gemelke2010rotating}, Raman couplings~\cite{LinSynthMag2009,cooper2011optical}, or lattice modulation~\cite{tai2016microscopy,PhysRevLett.111.185301,Jotzu:2014aa}, at the cost of reduced energy scales and challenges in state preparation~\cite{he2017realizing}. For materials made of light, synthetic magnetic fields have been realized across the electromagnetic spectrum, from the optical~\cite{HafeziM.:2013aa, Schine2016, Rechtsman:2013aa} domain, to microwave~\cite{Wang:2009aa} and even RF~\cite{PhysRevX.5.021031} photons. Because photons do not naturally interact with one another, the principal challenge is to realize a topological meta-material which is compatible with strong interactions. In the optical domain, Rydberg electromagnetically induced transparency is a possibility~\cite{peyronel2012quantum}; in the microwave, circuit quantum electrodynamics (cQED) tools offer a viable solution~\cite{wallraff2004strong,roushan2016chiral}.

In this work we engineer a synthetic time-reversal symmetry breaking magnetic field for microwave photons in a square lattice, where the magnetic length is twice the lattice vector. Importantly, we employ seamless 3D microwave cavities all machined from a single block of aluminum, so our meta-material is scalable and directly compatible with the cQED toolbox~\cite{anderson2016engineering} for entering the fractional Chern regime, as it is composed only of Aluminum for the cavities, plus Yttrium-Iron-Garnet (YIG) spheres and Neodymium magnets to produce the synthetic magnetic field. Circumnavigating a single plaquette induces a $\frac{\pi}{2}$ geometric phase, making the material equivalent to a quarter flux Hofstadter model.


In Section ~\ref{Sec:ExpSetup}, we describe and characterize the essential components of the Chern insulator, and explain how these elements are combined to realize a quarter flux Hofstadter model. Section ~\ref{Sec:FreqDomain} investigates the spectral properties of the realized model, observing four bulk bands and topologically protected edge channels living within the gaps between the top and bottom bands; using our single-site spatial resolution we are able to directly measure the dispersion of the edge channels. In Section ~\ref{Sec:Dyn} we measure edge transport around the Chern insulator, observing chiral, backscatter free dynamics. In Section ~\ref{Sec:Junction}, we demonstrate that the system may be reconfigured as a microwave tunnel-junction, and explore the dynamics across the tunnel-barrier, and Section ~\ref{Sec:Outlook} concludes.

\section{Experimental Setup}
\label{Sec:ExpSetup}

\begin{figure}
\includegraphics[clip=true, totalheight=7.2 cm]{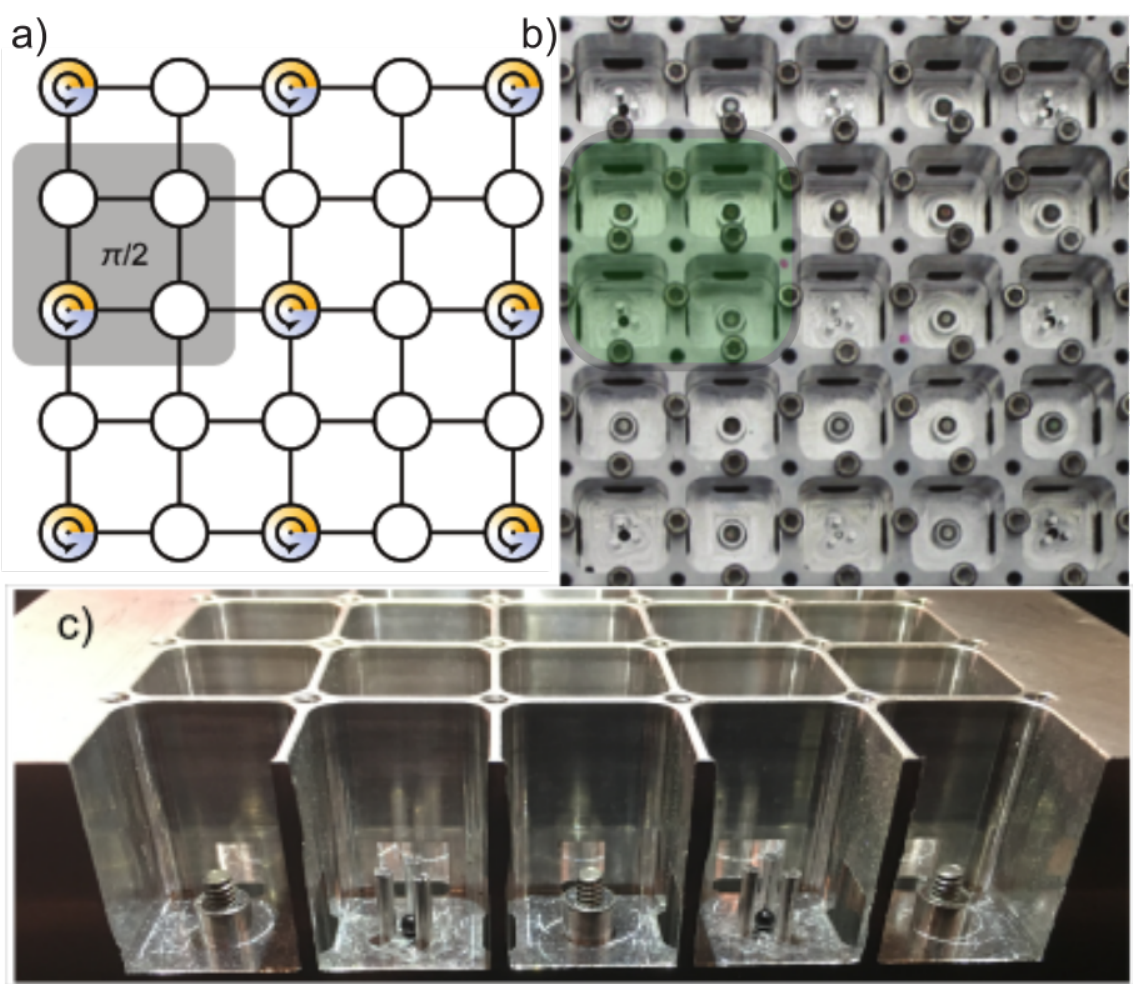}
\caption{Schematic and photograph of microwave Chern insulator lattice. \textbf{(a)} Schematic of the cavity layout. The white circles denote cavities that do not shift the phase of a photon passing through them, while the cavities with the arrows shift the phase depending on the physical angle between the couplers. This layout guarantees phase of $\frac{\pi}{2}$ per plaquette.
\textbf{(b)}Photograph of a 5x5 section of the lattice measured in the data presented in the paper. The lattice sites are each tuned to $9.560$ GHz $\pm 1$ MHz, and coupled evanescently tunnel coupled with hopping rate $30$ MHz $\pm 1$ MHz. The typical resonator quality factor for fundamental cavities is $Q=3000$ and for the YIG cavities is $Q=1500$. The lattice spacing is 1.96 cm, resulting in a total edge-to-edge (including outer walls) lattice dimension of 24.0 cm. \textbf{(c)} Side profile of a 5x5 lattice cut so that both types of cavities are visible. The 2nd and 4th cavities are the phase shifting YIG cavities, while the 1st, 3rd and 5th cavities are the fundamental cavities. The couplers are visible in between the cavities as gaps in the cavity walls. They are milled from the opposite side of the lattice as the cavities. \label{fig:combinedlatticepics}}
\end{figure}

To implement a quarter-flux Hofstadter lattice, we build upon the design described in ~\cite{anderson2016engineering}. The central innovation of this work is that while the Peierl's phase which encodes the synthetic gauge field is typically encoded in the tunnel coupling between lattice sites, in practice the phase can arise from the spatial structure of the sites themselves rather than manipulation of the tunnel couplers ~\cite{roushan2016chiral, PhysRevA.82.043811}. Engineering the on-site Wannier function of every fourth lattice site to exhibit a $2\pi$ phase winding ensures that every plaquette contains a phase-engineered site, as shown in Fig. ~\ref{fig:combinedlatticepics}a, and thereby induces a flux per plaquette of $\alpha=\frac{1}{4}$.

The metamaterial is milled into a block of aluminum (see Fig.~\ref{fig:combinedlatticepics}b) and is composed of three components: (i) fundamental mode cavities; (ii) chiral cavities; and (iii) evanescent couplers. The cavity design (Fig. \ref{fig:combinedlatticepics}c) enables machining of the structure from only two sides, and completely removes seam-loss as a source of Q reduction~\cite{Reagor-QuantumMemory}. Lattice sites are either single post coaxial resonators oscillating in their fundamental mode with a spatially uniform phase profile, or are three post coaxial resonators with a chiral phase profile designed to produce a synthetic gauge field. The final ingredient is the couplers, which induce the tunneling term in the Hofstadter Hamiltonian (Fig. \ref{fig:combinedlatticepics}c). The chiral cavities are designed so that a photon tunneling into and subsequently out of the cavity acquires a phase equal to the angle between input and output arms. When the lattice is arranged as shown in Fig. \ref{fig:combinedlatticepics}, a photon traveling in the smallest closed loop (plaquette) acquires a $\frac{\pi}{2}$ geometric phase.

\begin{figure}
\includegraphics{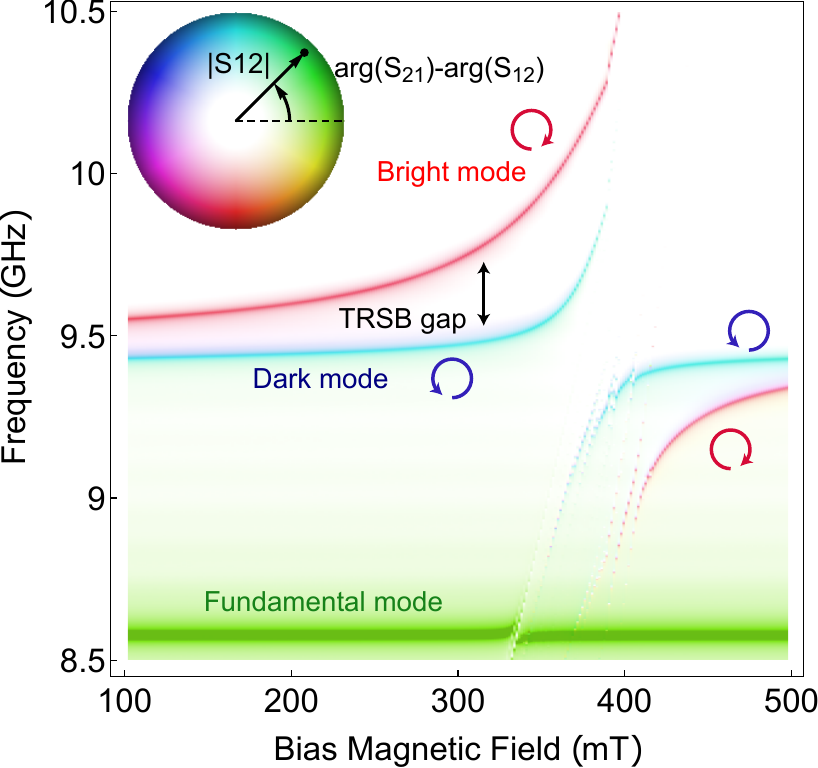}
\caption{$S_{12}$-$S_{21}$ Transmission between two antennas $45^\circ$ apart in a single YIG cavity as a function of magnetic field. There is stronger coupling between the cavity mode with the same chirality as the YIG sphere. The largest frequency splitting is $446$ MHz. The color indicates the phase shift a photon acquires in transmission. After subtracting the two different directions of transmission to eliminate phase noise from cables and impedance mismatches, one chiral mode has a phase shift of $90^\circ$ while the other has a phase shift of $-90^\circ$, indicating the modes are of opposite chirality. \label{fig:YigSpectroscopy}}
\end{figure}

In the fundamental mode cavities, a single post protrudes into an otherwise empty rectangular box. The length of the post sets the frequency of the resonator's fundamental mode to $9.560$ GHz (to within $1$ MHz), with the next mode at approximately at twice this frequency (see SI section \ref{SI-modes}). The side lengths of the box ensures that its cutoff frequency is higher than the post mode, resulting in localization of the post mode even without a lid on the resonator, with a mode Q determined by the length of the cylinder, until surface losses dominate.

The second type of cavity exhibits modes whose phase depends on the location in the cavity. This chiral cavity is structurally similar to the fundamental coaxial cavity but with three equal length posts arranged in an equilateral triangle at the center of the cavity, instead of a single post (Fig. \ref{fig:combinedlatticepics}c). These three closely spaced posts constitute three coupled degenerate resonators, and hence behave as a three-site tight-binding model with periodic boundary conditions. The result is one (quasi-momentum $q=0$) mode in which all posts oscillate with the same phase, and two degenerate modes at a higher frequency ($1$ GHz higher in this lattice), at $q=\pm\frac{2\pi}{3}$; the electric charge accumulation in these latter two modes travels from post-to-post clockwise or anti-clockwise, respectively.

To break the time-reversal symmetry of the lattice and thereby induce a chirality in the system, it is essential that only one of the two degenerate modes at $|q|=\frac{2\pi}{3}$ couples to the lattice bands. To achieve this, a $1$mm YIG sphere is inserted between the three posts (Fig. ~\ref{fig:combinedlatticepics}c). When a DC magnetic field $B_{DC}$ is applied to the YIG, it behaves as a macroscopic electron spin whose the magnetic moment precesses at the Larmor frequency $\omega_l=\mu_B B_{DC}$, and with a handedness set by the direction of the magnetic field ($\mu_B=28$MHz/mT is the Bohr Magneton). When the YIG sphere is installed between the three posts of the cavity, where the microwave magnetic field is strongest, the precessing magnetic moment couples strongly to cavity mode which co-rotates ($q=\frac{2\pi}{3}$), and weakly to the one which counter-rotates ($q=-\frac{2\pi}{3}$). Figure \ref{fig:YigSpectroscopy} shows the observed behavior of the $q=\pm\frac{2\pi}{3}$ modes as the DC magnetic field is varied, tuning the YIG frequency through the bare $|q|=\frac{2\pi}{3}$ mode frequency: the YIG induces a large avoided crossing with the co-rotating ``bright'' mode, a smaller asymmetry-induced avoided crossing in the counter-rotating ``dark mode''. Fine adjustment of the magnetic field strength can further be used to tune the frequency of the YIG cavities.

To probe the spatial structure of the chiral modes, we insert two antennae into one of the chiral cavities, separated by 45$^\circ$. Figure \ref{fig:YigSpectroscopy} shows the observed phase difference between exciting one and measuring the other, and the reverse, indicating that the ``dark'' mode exhibits a phase difference of 90$^\circ$, while the ``bright'' mode exhibits a difference of -90$^\circ$, consistent with their opposite chiralities.

To build a T-broken model, our engineered Hamiltonian must employ \emph{only one} of these two chiral modes; accordingly, the ``dark''-``bright'' mode splitting of $350-450$ MHz (depending on $B_{DC}$) sets the spectral domain into which the engineered Chern-band structure must fit. While the ``dark'' mode is the better choice in a superconducting resonator, YIG loss is much larger than the bare resonator loss, in our case technical concerns make the ``bright'' mode preferable for lattice engineering (see SI \ref{SI-modes}).

The lattice-sites are tunnel coupled by milling out slots between them; because the lowest mode of these slots is above the cutoff of the lattice, these slots couple sites together without inducing radiative loss. The width and depth of the couplers set the tunneling energy, which we tune with a screw (see SI \ref{SI-modes}) to $30$ MHz with precision $\pm 1$ MHz, in accordance with the requirement that the total band structure be narrower than the minimum $350$ MHz ``dark''-``bright'' splitting.

\par
\section{Spectral Properties of a Microwave Chern Insulator}
\label{Sec:FreqDomain}
\begin{figure}
\includegraphics[width=3.5in]{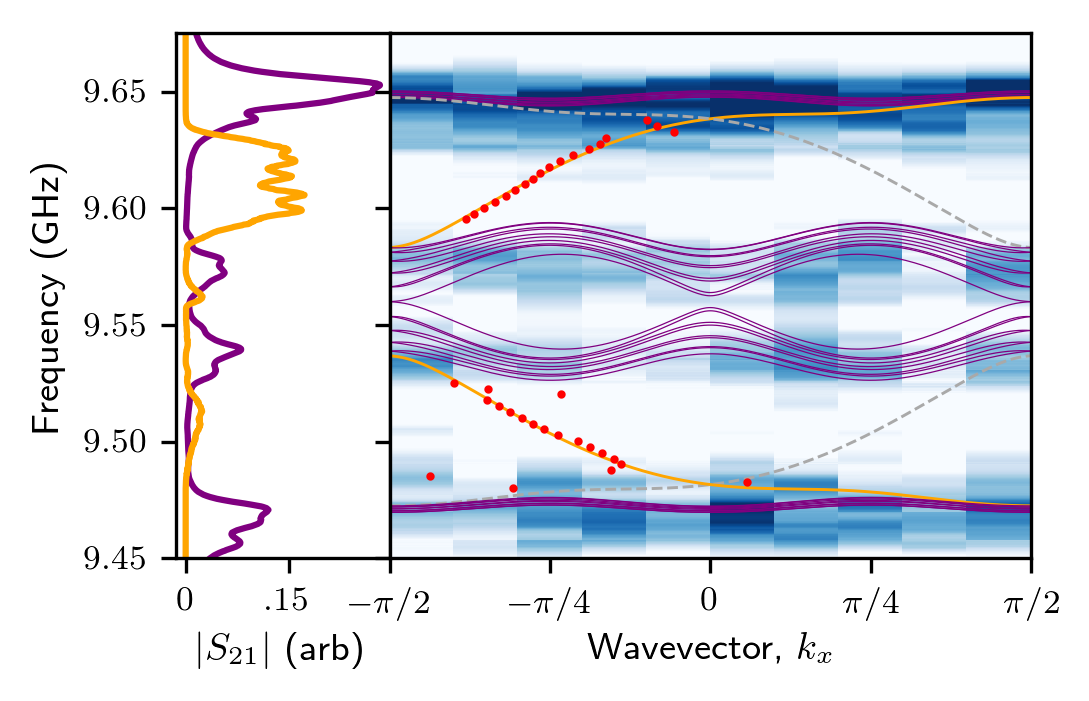}
\caption{\label{fig:BulkBulkandEdgeEdge} \textbf{(a)} Measured transmission spectrum between two bulk (purple) and edge (orange) sites. The purple trace is the transmission between adjacent cavities in the bulk of the lattice ((5,6) and (6,6) defined from the upper-left lattice corner), while the orange trace is the transmission between cavities on the edge of the lattice (sites (1,1) and (1,11)). The differential response between upper and lower gaps arises from the distance difference for clockwise and anti-clockwise edge propagation between probe and measurement sites.  \textbf{(b)} Projected band structure of both the bulk (blue/white density plot, with spatial Fourier limited resolution) and edge (red points) of the system, compared with theory for a $\alpha=1/4$ Hofstadter strip (purple/orange/gray-dashed). The bulk data results from a site-by-site measurement of the system response. The dispersion of the edge channel (explained in the SI) is extracted from the measured cavity-to-cavity phase shift. Within the bulk bands, this phase is sensitive to disorder and overall geometry, resulting in a near-random signal which we omit from the plot. The bulk-bulk transmission exhibits four distinct bands which arise from the four site magnetic unit cell. Gaps are apparent between the first and second bands as well as the third and fourth bands. The magnetic unit cell has two sites in each direction, compressing the Brillioun zone to $\left[-\frac{\pi}{2},\frac{\pi}{2}\right]$.}
\end{figure}

A defining characteristic of a topologically nontrivial band structure is an insulating bulk and conducting edges~\cite{RevModPhys.82.3045}. To demonstrate that our microwave lattice exhibits these properties, we probe it spectroscopically by placing a dipole antenna into each cavity. Figure \ref{fig:BulkBulkandEdgeEdge}a shows a typical transmission spectrum between bulk lattice sites. We observe energy gaps in the bulk response, within which the edge-localized channels reside, consistent with the computed band structure~\cite{anderson2016engineering} shown in Fig. \ref{fig:BulkBulkandEdgeEdge}b. The chirality of the edge channels results in an asymmetric edge-edge response in the band gaps: the edge modes in the lower and upper gaps have opposite group velocities and finite damping, so the accrued decay is smaller when the excitation travels the ``short way'' versus the ``long way''.

\begin{figure}
\includegraphics[width=3.5in]{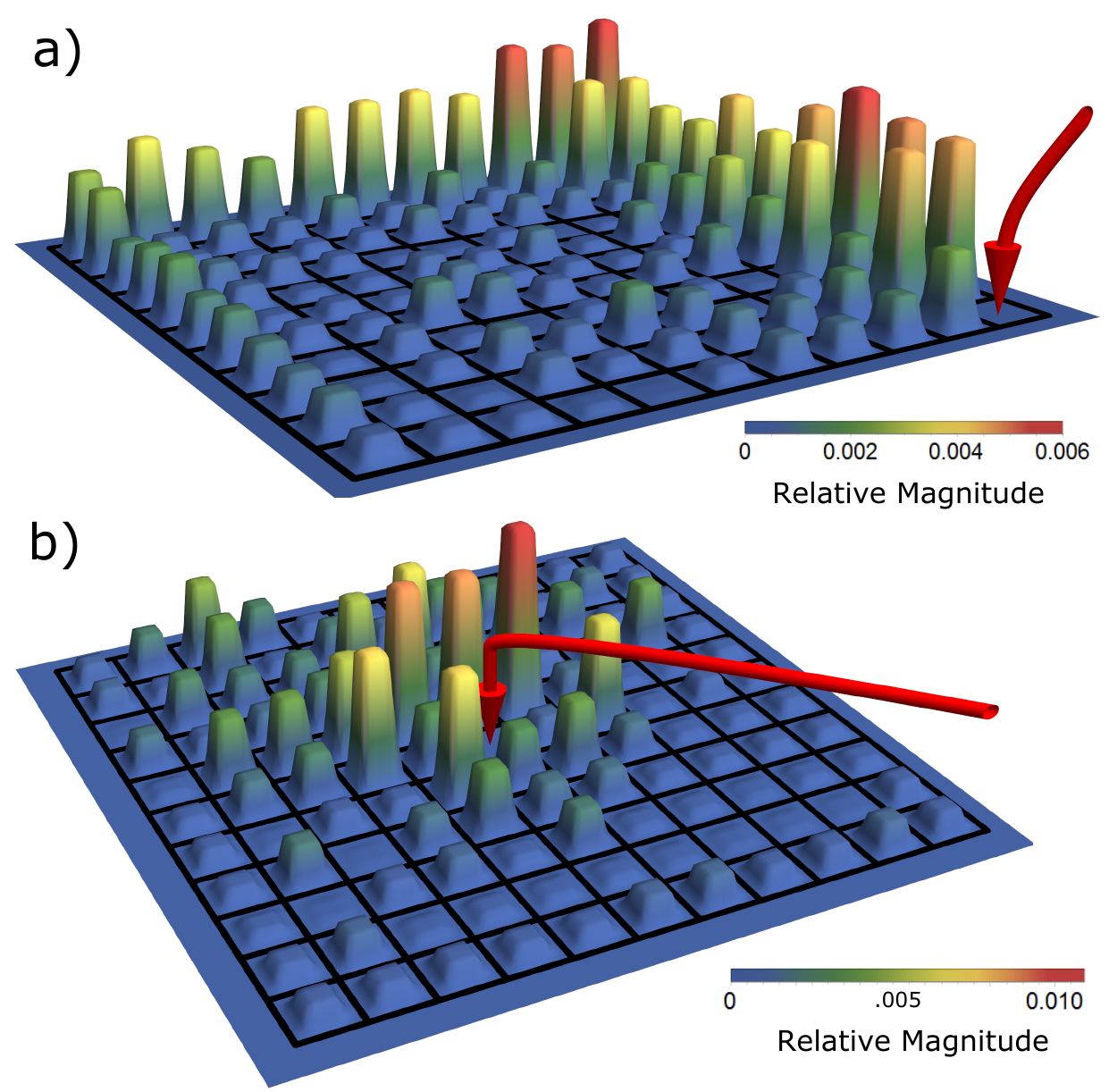}
\caption{\label{fig:ContandPulse}  \textbf{(a)} The lattice's response to an edge-excitation (red arrow), at the a frequency of $9.622$ GHz which is within the band gap. The presence of an edge channel at this frequency results in a delocalized chiral response along the system edge, decaying due to the finite resonator Q's. \textbf{(b)} The response of the lattice when a bulk site (red arrow) is excited continuously at a frequency of $9.569$ GHz within the upper band. The absence of bulk modes at the excitation frequency results in exponential localization of the response.}
\end{figure}

When the system is excited in the bulk within the bulk band gap, we observe a localized response as shown in Fig. \ref{fig:ContandPulse}a, resulting from vanishing density of states in the bulk at energies within the band gap. On the other hand, when the system is excited on its edge within the bulk band gap, we observe the delocalized response shown in Fig. \ref{fig:ContandPulse}b), resulting from the presence of a chiral edge channel within the bulk energy gap. Decay in the lattice allows us to observe the chirality of the edge mode in the steady state response; at $9.6$ GHz, the channel travels counterclockwise, as anticipated from the band structure in Fig. ~\ref{fig:BulkBulkandEdgeEdge}b. By taking a Fourier transform of the spatially-resolved  (complex) transmission, we are able to reconstruct the bulk band structure of the lattice. For simplicity, we project out a transverse spatial coordinate, and plot the resulting 1D band structure in blue in figure \ref{fig:BulkBulkandEdgeEdge}b.

\begin{figure}
\includegraphics[width=3.5in]{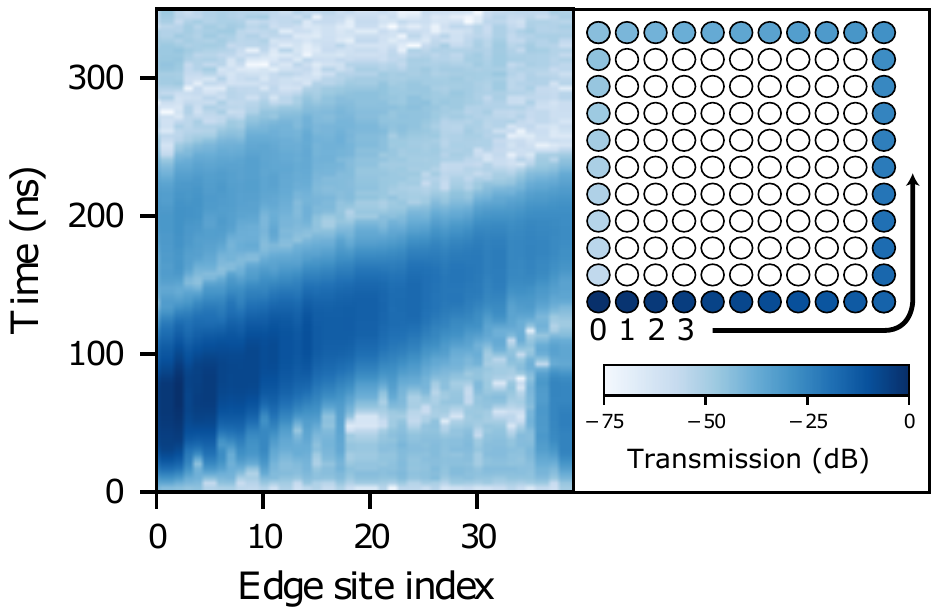}
\caption{Spatio-temporal response of the edge cavities to a $50$ ns pulse centered on $9.6$ GHz. The x-axis is the indexed cavity number of each of the 40 edge sites, and the y-axis is time. The brightness reflects the relative normalized transmission between the excited cavity and the measured cavity. The chirality of the edge channel is reflected in the unidirectional travel of the pulse. The weak stationary response results from a small Fourier-broadened excitation of bulk modes.\label{fig:PulseNoWall} }
\end{figure}

The site-resolved probes accessible in our system enable us to directly access, for the first time, the dispersion of the edge channel, which we achieve by probing the system edge within the bulk energy gap and measuring the phase accrued per lattice site as a function of frequency (See SI \ref{Dispersion}). Figure ~\ref{fig:BulkBulkandEdgeEdge}b shows the observed dispersion in red, in good agreement with the theoretical prediction shown in gray.

\section{Dynamics of Microwave Chern Insulator}
\label{Sec:Dyn}
A major advantage of using microwave photons is that time-resolved edge-transport measurements are possible. Instead of exciting the lattice with a CW signal from a network analyzer, we can also apply a pulse to single site on the edge and observe its chiral propagation. In what follows, we excite the center of the upper bandgap ($9.6$ GHz) with the shortest Gaussian pulse not Fourier broadened into the nearby bulk bands ($75$ ns), at site (1,1). We then measure the response at each site as the pulse pulse propagates. Figure \ref{fig:PulseNoWall} shows the response of the system edge; the pulse travels in one direction with a well-defined and constant velocity, exhibiting no backscattering due to the protection provided by the chirality of the system. The group velocity is measured to be $0.32 \pm .04$ sites/ns, consistent with the measured dispersion $\frac{\partial \omega}{\partial k}$ at $9.6$ GHz ($0.328 \pm .002$ sites/ns). Weak Fourier broadening into the bulk bands is also apparent in the data as a small excitation fraction over all sites that does not propagate. See the supplemental information \ref{Dispersion} for a video of the pulse propagating through the lattice.

\section{Photonic Tunnel Junction}
\label{Sec:Junction}

Looking forward to exploration of strongly interacting topological phases~\cite{anderson2016engineering}, it will be essential to fabricate metamaterial structures which operate as spatial interferometers~\cite{mcclure2012fabry,an2014fabrication}. Such devices afford direct sensitivity to the charge and statistics of edge excitations through response of the interference fringe to magnetic flux and enclosed anyons, respectively. 

As a first step towards this objective, we harness the extraordinary flexibility of our platform to realize a photonic tunnel junction, analogous to half of an solid-state edge channel interferometer~\cite{an2011braiding}. We realize the  tunnel junction by detuning a single column of lattice sites in our 11x11 sample, leaving only a single site in the center of column at its original frequency; this produces a ``wall'' between two subsamples, with a narrow gap through which photons may tunnel.

\begin{figure}
\includegraphics[width=3.5in]{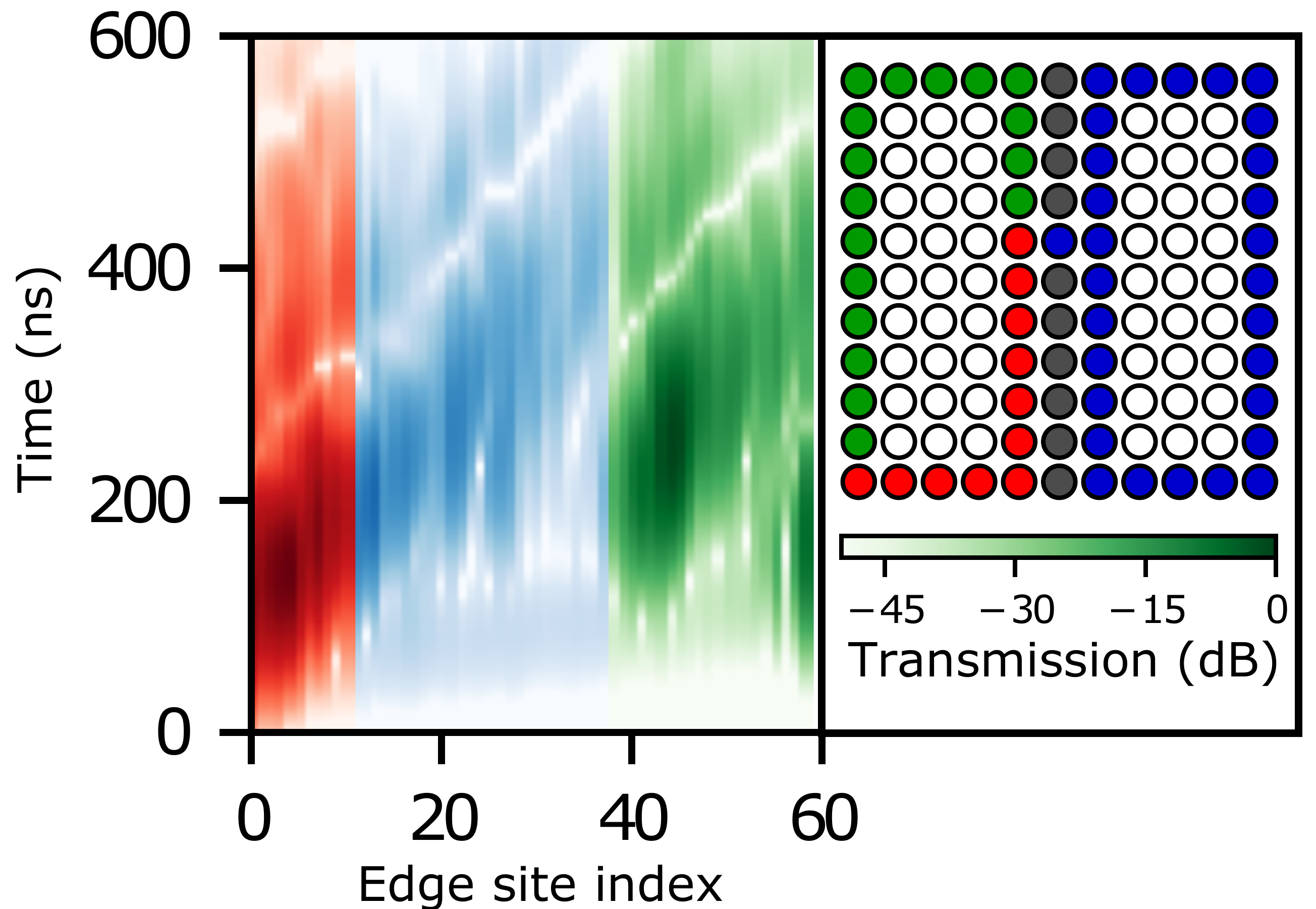}
\caption{\label{fig:Wall Measurement} A wall is built into the lattice by detuning all but one cavities in the sixth row, separating the lattice into two 11x5 lattices connected by one cavity. This effectively makes a beam-splitter for running wave edge modes, as shown schematically at \textbf{right}. When the lattice is pulsed at $9.6$ Ghz (in the top band gap), the response is shown in the figure on the left. The pulse splits when it reaches the gap in the wall (end of the red), transmitting most of the pulse to the green part of the lattice and some of the pulse into the blue, unexcited 11x5 sublattice.}
\end{figure}

Figure \ref{fig:Wall Measurement} shows this arrangement; a propagating edge excitation may either tunnel across the gap and stay in the original ring in which it was traveling, or continue along the edge into the neighboring ring, akin to a chiral edge beam-splitter. Results are shown in Fig.~\ref{fig:Wall Measurement} for an excitation starting at cavity site (1,1). It bears mentioning that while a fraction of the excitation remains in the original ring and a fraction hops to the other ring, none is back-scattered, illustrating robustness of the edge channel to disorder. The added disorder brings edge channels from both sides of the lattice together, opening a path for photons from one side to enter a backwards edge mode of the other side of lattice. However, the photons only travel with one chirality, as shown in Figure \ref{fig:Wall Measurement}.\\

\section{Outlook}
\label{Sec:Outlook}
We have demonstrated a complete toolbox for the development of low-loss topological microwave lattices, and harnessed this toolbox to realize a quarter flux Hofstadter model. The resulting synthetic material is be probed site-by-site, revealing an insulating bulk and topologically protected chiral edge channels. We showcase the flexibility of the approach by reconfiguring the lattice to act as a tunnel junction, pointing the way to anyon interferometry once circuit quantum electrodynamics tools provide interactions between lattice photons.

Looking ahead, the next step is to marry these microwave resonator arrays with the tools of circuit quantum electrodynamics ~\cite{wallraff2004strong}, thereby inducing on-site interactions. These interactions correspond to a Hubbard $U$ in the Hofstadter-Hubbard model, immediately enabling studies of fractional quantum hall phases of interaction photons~\cite{anderson2016engineering}. Because the demonstrated lattice is already low-loss at room temperature ($\sim 3.5$ MHz linewidth), the typical transmon qubit anharmonicity of $200$ MHz ~\cite{koch2007charge} is sufficient to induce strong correlations between lattice photons. To prepare the photons in low entropy phases of the resulting models, it will be crucial to harness state-of-the-art theoretical tools to populate these models near their manybody ground states. For small systems, this will rely upon spectroscopically resolved excitation of manybody states~\cite{umucalilar2013many}, while for larger systems engineered dissipation~\cite{kapit2014induced, hafezi2015chemical, ma2017autonomous, biella2017phase} will allow for preparation of incompressible phases. In sum, the platform opens many exciting prospects at the interface of topology, many-body physics, quantum optics, and dissipation.

\section{Acknowledgements}
This work was primarily supported by the University of Chicago Materials Research Science and Engineering Center, which is funded by National Science Foundation under award number DMR-1420709. This work was supported by ARO grant W911NF-15-1-0397; D.S. acknowledges support from the David and Lucile Packard Foundation; B.M.A. and R.M. acknowledges support from a MRSEC-funded Kadanoff-Rice Postdoctoral Research Fellowship; C.O. is supported by the NSF GRFP. The authors thank D. Angelakis, W. DeGottardi, M. Hafezi, and M. Levin for helpful discussions.

\bibliographystyle{unsrtnat}
\bibliography{Quarter-FluxHofstadter}

\clearpage
\renewcommand{\thefigure}{S\arabic{figure}}
\setcounter{figure}{0}
\renewcommand\theequation{S\arabic{equation}}
\setcounter{equation}{0}

\section{Supplementary Information}
\subsection{Engineering the Modes in Coaxial Microwave Cavities}\label{SI-modes}

\begin{figure*}
\subfloat[\label{fig:Fun}]{\includegraphics[width=0.25\textwidth]{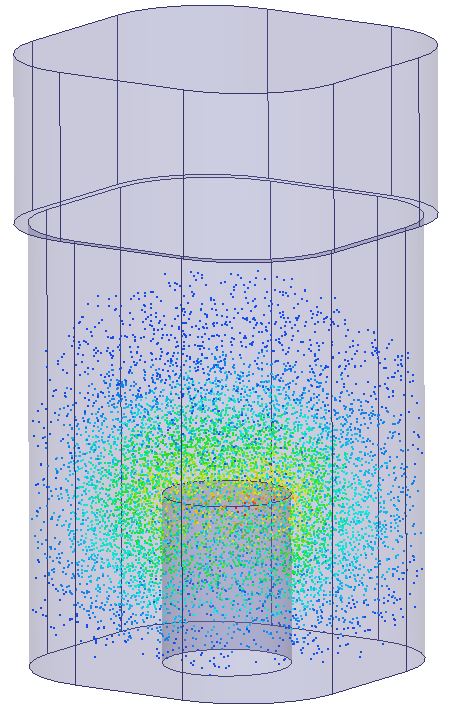}}\hspace{0.15ex}
\subfloat[\label{fig:YigE}] {\includegraphics[width=0.33\textwidth]{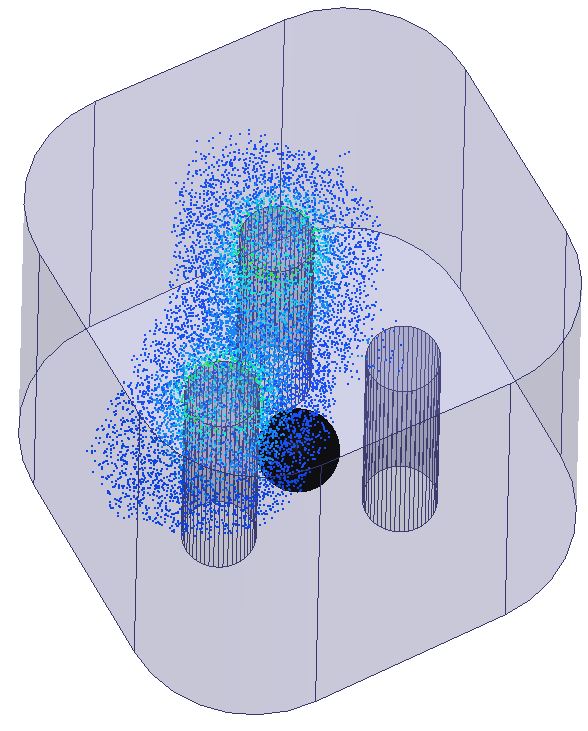}}
\subfloat[\label{fig:YigB}] {\includegraphics[width=0.35\textwidth]{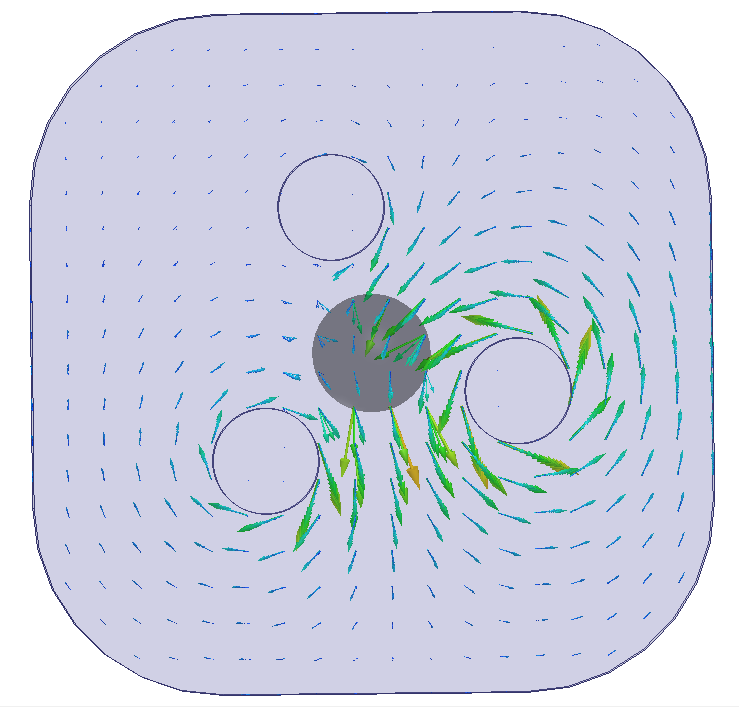}}
\caption{\textbf{(a)} Fundamental mode cavity electric field. The mode is localized at the bottom of the cavity.  \textbf{(b)} Electric field simulation of one of the chiral YIG cavity modes. A video that shows the rotation as a function of time is available at https://youtu.be/gHoyFxkw9iQ. \textbf{(c)} Magnetic field of one of the chiral YIG cavity modes viewed from the top. The field at the center of the cavity rotates in time, coupling to the magnetic moment of the YIG.} \label{CavitySims}
\end{figure*}

In our prior proposal~\cite{anderson2016engineering}, cylindrical cavities were used to create the degenerate mode structure necessary from which the chiral modes are constructed.  While the cylindrical cavity, supports the required modes, it has two technical limitations.  First, the diameter must be of order $\lambda$ which makes the lattices physically large.  More importantly, it is difficult to construct cylindrical cavities without a seam, introducing associated losses\cite{brecht2015demonstration}.  In typical 3D qubit experiments\cite{paik2011observation}, the seam is cut such that no current crosses it significantly reducing loss.  However, for chiral modes this cannot be done, and quality factor is limited by this loss mechanism even at room temperature, especially for larger lattices.  A significant innovation in this work is to use seamless cavities engineered to have appropriate mode structure.  This geometry is both compact, and has no seam loss, making it ideal for current and future studies.

Cavities with a single post support two varieties of microwave modes:

\begin{itemize}
\item{A mode whose frequency is dependent primarily on the length of the post (Fig. \ref{fig:Fun}). This mode is analogous to a coaxial cable which is shorted at one end and open at the other, with the post acting as the coax center pin and the outer wall acting as the coax shield. In this quarter-wave resonator, the mode frequency is approximately four times the post length, and higher order modes are odd integer multiples of this fundamental frequency.} 
\item{Modes originating from the box in which the post resides. The lowest frequency of this type of mode is set by the two smallest dimensions of the box. We choose a cross-section that is small enough that the lowest frequency box mode is $1$ GHz above the fundamental mode of the coaxial resonator.}
\end{itemize}

As we reduce the cross section of the box, the surface-to-volume ratio increases, thereby increasing resistive surface losses, the dominant loss channel for room temperature resonators. Accordingly, we choose a radius that maximizes the Q of the post modes while ensuring the cutoff of the box modes is well above the frequency of the fundamental post mode (to avoid resonant/evanescent outcoupling of energy from the post mode to the outside world through the box modes). Since the frequency of the post mode is strongly dependent on the length of the post, an aluminum screw (low-loss dielectric can be used at low temperatures) threaded into a hole in the post enables fine control of the post mode frequency to $\leq0.01$\% ($1$ MHz).

The three-post cavities exhibit similar box modes, but with the addition of two more post modes (one mode per post). In order to break time reversal symmetry we first create two degenerate modes with opposite chirality (thereby preserving time-reversal symmetry) that maintain the loss properties of reentrant seamless cavities. (Single post cavities exhibit an excited manifold composed of two time-reversal symmetric modes, but these are at approximately twice the frequency of the fundamental mode and the position of the maximum magnetic field moves as a function of time, making these cavities incompatible with coupling to a ferrite.) The additional post modes couple to one another through the box modes. In the case of three posts, the spectrum of fundamental post modes resolves into a $q=0$ mode where all posts resonate in phase, and two degenerate higher energy modes with $q=\pm\frac{2\pi}{3}$, where the field's maximum hops from post to post either clockwise or counterclockwise. The mode structure is shown in Fig. \ref{fig:YigE} (electric field) and in Fig. \ref{fig:YigB} (magnetic field) and animations are available in the supplemental materials. The magnetic field is not maximal at cavity center, but exhibits sufficient concentration to couple strongly to the ferrite (YIG sphere); at resonance, the bright mode to ferrite coupling reaches $1.4$ GHz.

In order to break time reversal symmetry, we break the degeneracy between the two circular modes in the three post cavities. This gives a linear angular dependence to the phase of the cavity mode, since a rotating cavity modes acquires phase as it rotates through the cavity. For these cavity modes, a photon traveling a full rotation around the cavity shifts it phase by $2\pi$. To break the degeneracy, we couple the rotating modes of the cavity to a ferrite material called YIG (yttrium iron garnet) that has its own chiral modes. When a magnetic field is applied to YIG, it behaves like coherent electron spins. The magnetic moment precesses at a frequency $f=\gamma B$, where $\gamma$ is the gyromagnetic ratio $28$ GHz/T and $B$ is the DC field strength. This precession has the same chirality as one of the cavity modes and the opposite of the other mode. Due to what amounts to a rotating wave approximation in real space, the YIG couples much more strongly to the cavity mode precessing with the same frequency. In fig \ref{fig:YigSpectroscopy}, we show the magnetic field dependence of the frequency of the circular modes. One of the modes shifts in frequency more than the other as we tune the magnetic field, giving a maximum frequency separation in the cavities of ~400 MHz. The phase shift through the bright mode is the same and equal to two times the angle between the antenna (the factor of 2 comes from measuring S12-S21 to eliminate cable and connector phase), while the phase shift through the dark mode is 2Pi minus the phase shift in the bright modes since this mode is orbiting with opposite chirality. The uniform oscillating mode is also shown on this plot at the lowest frequency. The phase shift through this mode is $0$ since it has no spatial dependence on phase. 

The dark mode does interact to some degree with the YIG spheres in these types of cavities (as opposed to perfectly circular cylindrical cavities).  This is primarily because the modes of the resonators are not perfectly circularly polarized and because the bias field is not perfectly homogeneous.  Additionally, the coupling to the YIG sphere is quite strong, making these effects easily observable as additional avoided crossings in Fig.~\ref{fig:YigSpectroscopy}. At low temperatures we will couple the lattice to the dark mode since it will hybridize less with the YIG modes (cavity modes are higher Q when the cavity is superconducting). At room temperature it is convenient to couple to the bright, higher frequency YIG mode since it not only has similar Q to the dark mode, but also only has the one dark mode nearby in frequency. Also, since it is higher frequency, it allows the fundamental cavities to have less screw length protruding into the cavity and thus somewhat better Q's.

The couplers can be thought of as higher frequency resonators between neighboring cavities that couple the cavities with a strength $\frac{g^2}{\Delta}$, where $g$ is the direct coupling of the cavities to the coupler and $\Delta$ is the detuning. Thus one way to increase the coupling between cavities is to bring the posts of the cavities closer together which increases $g$. Similarly, any method that concentrates the mode of the coupler closer to the posts will increase the coupling. To increase $\Delta$, we can either increase the size of the coupler or add a post in the coupler so that the coupler has its own post mode. By carefully selecting the length of this post, we can keep the couplers off-resonant to the lattice, but decrease $\Delta$ significantly for higher coupling. We use a tapped screw as the post so that we can tune the coupling between lattice sites. The screws allow the frequency of the coupling to be tuned from $20$ MHz to $100$ MHz, though to keep the band structure of our lattice comfortably within the $400$ MHz splitting between the chiral YIG cavity mode (so that the YIG mode with opposite chirality does not hybridize with any lattice modes), we tuned the coupling in this paper to $30$ MHz.

\subsection{Measuring the Lattice}\label{Lattice Measurements}

\begin{figure*}
\includegraphics[clip=true, totalheight=9 cm]{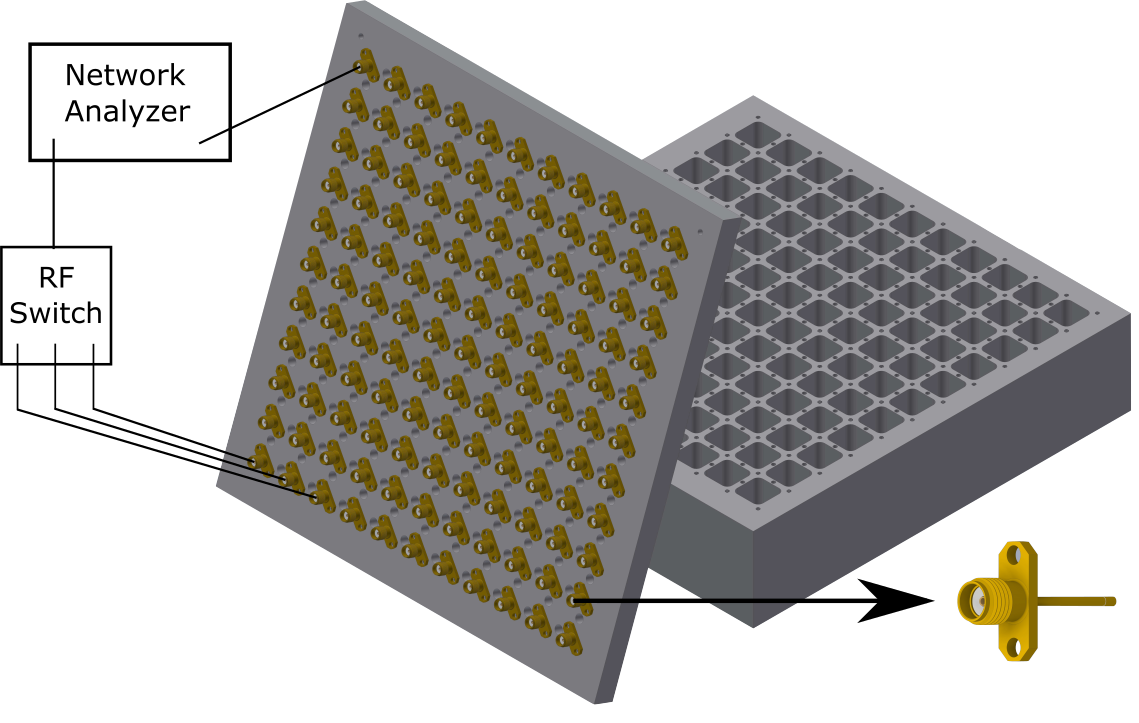}
\caption{\label{fig:LatticeWithLId} 
The lattice is shown on the right. A lid for the lattice is made so that one of the gold-colored antennas is connected to every lattice site. These antennas are connected through an RF switch to the network analyzer so that arbitrary transmission between pairs of sites can be measured.}
\end{figure*}

Every cavity has a microwave antenna weakly coupled to it from the top, so that the antenna does not add significant loss or shift the frequency of the resonator. The length of all the antennas are kept the same so that the coupling to each cavity is the same. Each antenna is then connected to a vector network analyzer through a switch network so that we can measure the transmission between any two cavities. We can measure reflection off any site as well, though reflection measurements are more sensitive to impedance mismatches in the cables and switches.  Effectively, allows us to perform measurements akin to an scanning tunneling microscope, for microwave metamaterials.

Using the same network of antennas, we can pulse the lattice and measure the response as a function of time. To create the pulse, we mix a 9.6 GHz sine wave signal with a $75$ ns long Gaussian pulse ($50$ ns for the wall experiment). The pulse must be short enough in the time domain so that it does not interfere with itself, but long enough so that the pulse is not so wide in frequency space that it strongly excites the bulk bands. We use a shorter pulse for the wall data since the pulse takes less time to come back to the originally excited cavity. To measure the pulse, we first use an IQ mixer to mix the signal coming out of the measured cavity with the $9.6$ GHz oscillator to make the signal near DC. We then measure the IQ output on a scope to get both the phase and the amplitude of the response as a function of time.

\subsection{Dispersion of the Edge Channel}\label{Dispersion}

\begin{figure}
\includegraphics[totalheight=5 cm]{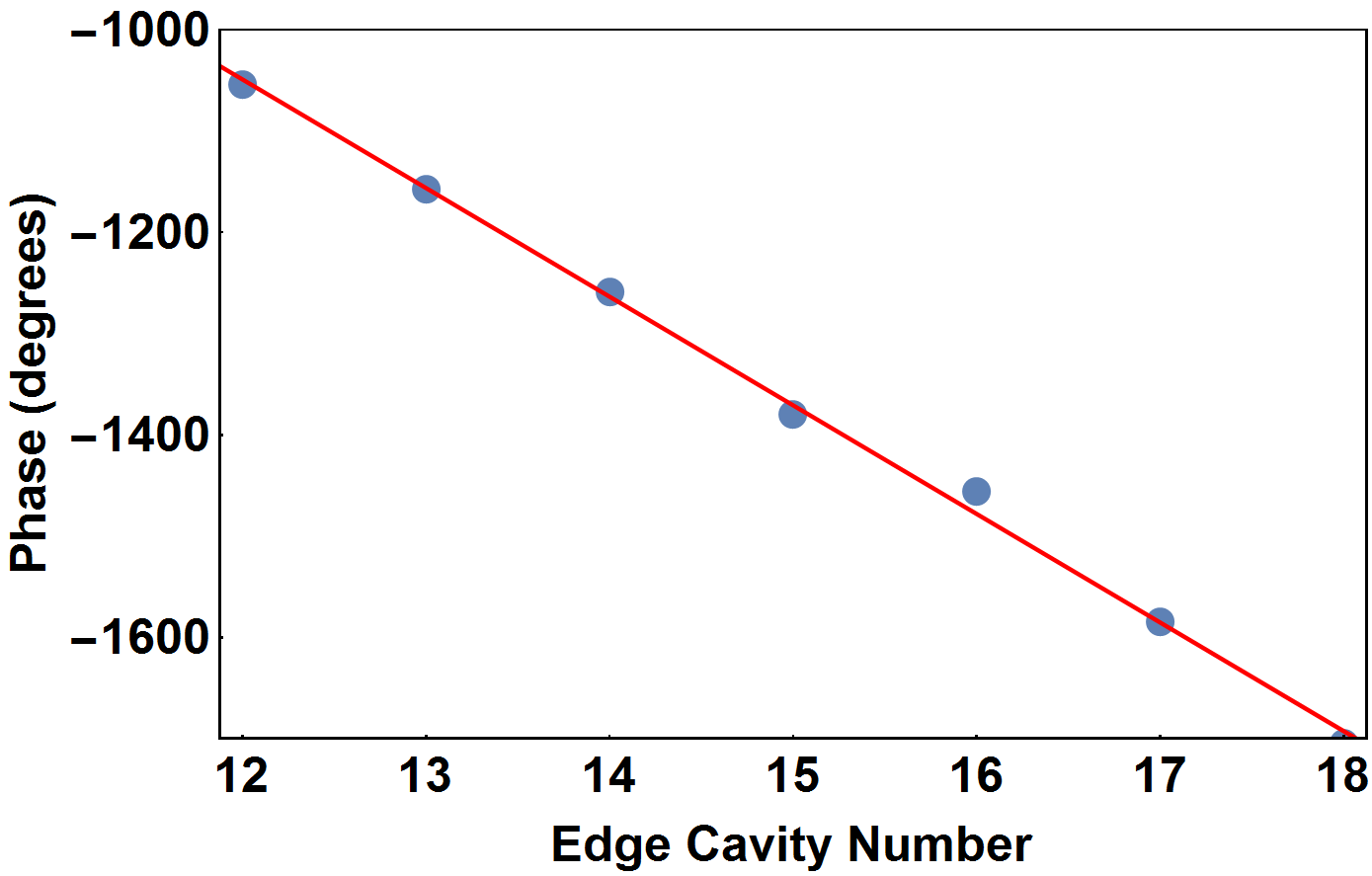}
\caption{\label{fig:PhaseUnwrapExample} We measure the difference in phase between an excited edge cavity and other edge cavities as a function of distance. The edge cavity number starts at 12 and ends at 18 so that only one edge is fit to a line, do avoid additional dispersion introduced by the corner. We choose the side furthest from the the excited cavity to minimize the effects of direct coupling. For the frequency shown here ($\omega=2\pi\times 9.61$ GHz) the lattice momentum is half the slope, or $-54\deg$ per cavity.}
\end{figure}

The dispersion an edge channel can be measured directly from the evolution of the phase response along an edge of the lattice, when the system is excited on an edge within the bulk band-gap. In a square lattice, the dispersion is constant along a side but changes near the corners, so we examine the phase response on non-corner sites along a single side of the lattice (i.e. sites (11,2) to (11,10)). Plotting the phase shift as function of distance between the excitation port and the measuring port yields a line with slope equal to the lattice momentum (see Fig. \ref{fig:PhaseUnwrapExample} for a sample data set). In our measurements, we subtract the S12 direction from the S21 direction in order to eliminate the phase shifting from cable length and connectors. This subtraction means that the slope of the line is actually twice the lattice momentum. The S12 excitation moves around the lattice the opposite direction as the S21 excitation, so they travel a different distance. When S21 is subtracted from S12, this discrepancy manifests itself as a constant offset proportional to the perimeter of the lattice; the slope of the phase vs distance plot is unaffected. The phase shift per cavity can be a significant fraction of $2\pi$, so phase unwrapping is necessary to recover the slope. Using this technique at a frequency within the bulk bands resuls in a near random signal since this phase at that frequency is sensitive to disorder and overall geometry,

A video of pulse propagation can be found in the online supplementary materials. In this video, the measured response is renormalized at each time before plotting, so that even after decay the pulse can be seen clearly.

\end{document}